\documentclass[conference,10pt]{IEEEtran}
\usepackage{epsfig,rotating,setspace,latexsym,amsmath,epsf,amssymb,amsfonts,bm,theorem,subfigure,epstopdf}
\usepackage{cite,authblk}
\usepackage{bbm}

\usepackage{algorithm}
\usepackage[noend]{algpseudocode}

\algrenewcommand\algorithmicforall{\textbf{foreach}}
\algrenewcommand\algorithmicindent{.8em}

\usepackage{color}
\usepackage{mathtools}

\IEEEoverridecommandlockouts
\allowdisplaybreaks

\begin{document}

\title{Who Should Google Scholar Update More Often?  \thanks{This work was supported by NSF Grants CCF 17-13977 and ECCS 18-07348.}}

\author{Melih Bastopcu \qquad Sennur Ulukus\\
	\normalsize Department of Electrical and Computer Engineering\\
	\normalsize University of Maryland, College Park, MD 20742\\
	\normalsize  \emph{bastopcu@umd.edu} \qquad \emph{ulukus@umd.edu}}

\maketitle

\begin{abstract}	
We consider a resource-constrained updater, such as Google Scholar, which wishes to update the citation records of a group of researchers, who have different mean citation rates (and optionally, different importance coefficients), in such a way to keep the overall citation index as up to date as possible. The updater is resource-constrained and cannot update citations of all researchers all the time. In particular, it is subject to a total update rate constraint that it needs to distribute among individual researchers. We use a metric similar to the age of information: the long-term average difference between the actual citation numbers and the citation numbers according to the latest updates. We show that, in order to minimize this difference metric, the updater should allocate its total update capacity to researchers proportional to the \emph{square roots} of their mean citation rates. That is, more prolific researchers should be updated more often, but there are diminishing returns due to the concavity of the square root function. More generally, our paper addresses the problem of optimal operation of a resource-constrained sampler that wishes to track multiple independent counting processes in a way that is as up to date as possible.
\end{abstract}
 
\section{Introduction}
\begin{figure}[t]
	\centering  \includegraphics[width=1\columnwidth]{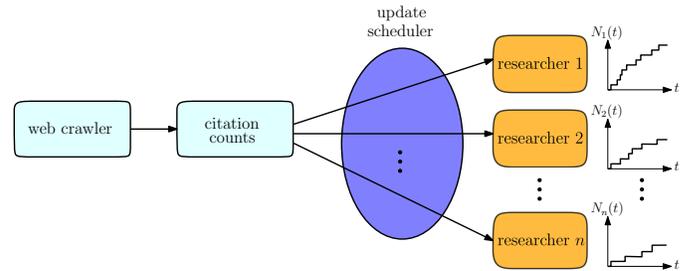}
	\caption{Web crawler finds and indexes scientific documents, from which citation counts are extracted upon examining their contents. Scheduler schedules updating citation counts of individual researchers based on their mean citations, and optionally, importance factors, subject to a total update rate.}
	\label{fig:model}
	\vspace{-0.5cm}
\end{figure}

Consider a citation index such as Google Scholar. As abstracted out in Fig.~\ref{fig:model}, Google crawls the web to find and index various items such as documents, images, videos, etc. Focusing on scientific documents, Google Scholar further examines the contents of these documents to extract out citation counts for indexed papers. Google Scholar then needs to update citation counts of individual researchers, which there are many. We model the citation count of each individual researcher as a counting process with a fixed mean, e.g., $\lambda_i$ for researcher $i$. Assuming that Google Scholar is resource-constrained, i.e., that it cannot update all researchers all the time, how should it prioritize updating researchers? If it can update only a fraction of all researchers, who should it update? Should it update researchers with higher mean citation rates more often as their citation counts are subject to larger change per unit time? Or should it update researchers with lower mean citation rates more often in order to capture rarer more informative changes?      

We view this problem with the lens of the recent literature on \emph{information freshness} quantified through the metric of age of information. Freshness, and age, of information have been studied in the context of web crawling \cite{Cho03, Brewington00, Azar18, Kolobov19a}, social networks \cite{Ioannidis09}, queueing networks \cite{ Kaul12a, Costa14, Bedewy16, He16a, Kam16b, Sun17a, Najm18b, Najm17, Soysal18, Soysal19}, caching systems \cite{Yates17b, Tang19}, remote estimation \cite{Wang19a, Sun17b, Sun18b, Chakravorty18}, energy harvesting systems \cite{Arafa17b, Arafa17a, Wu18, Arafa18b, Arafa18a, Arafa18f, Arafa19e, Farazi18, Leng19, Chen19}, fading wireless channels \cite{Bhat19, Ostman19}, scheduling in networks \cite{Nath17, Hsu18b, Kadota18a, Kosta17a, Bastopcu18, bastopcu_soft_updates_journal, Buyukates18c, Buyukates19b}, multi-hop multicast networks \cite{ Zhong17a, Buyukates18, Buyukates19, Buyukates18b}, lossless and lossy source coding \cite{Zhong16, Zhong18f, Mayekar18, partial_updates,   MelihBatu1}, computation-intensive systems \cite{Gong19, Buyukates19c, Zou19b, Arafa19a, Bastopcu19, Bastopcu20b}, vehicular, IoT and UAV systems \cite{ Elmagid18, Liu18}, reinforcement learning \cite{Ceran18, Beytur19, Elmagid19} and so on.  

We model the citation count of each researcher as a counting process with a given mean. In particular, we model citation arrivals for researcher $i$ as a Poisson counting process with rate $\lambda_i$. Optionally, we may further assign an importance factor to each researcher, based on their research field or citation count, but this is optional, and does not affect the structure of the results. If an importance coefficient is used, we denote it with $\mu_i$ for researcher $i$. Ideally the updater should update all researchers all the time, however, due to computational limitations, this may not be possible. We model the updater as a resource-constrained entity which has a total update capacity of $c$, which it should distribute among all researchers. We allocate an update rate $\rho_i$ for updating researcher $i$. These $\rho_i$ are collectively subject to the total system update capacity of $c$. We consider the cases of Poisson updates (i.e., updates with exponential inter-update times), deterministic updates, and synchronized updates. We determine the optimal update rates $\rho_i$ subject to the total update rate $c$ in a way to maximize the system freshness. In a broader sense, this problem is related to the problem of real-time timely estimation of signals, which have different change rates and importance factors, with the goal of finding the optimal individual sampling rates, under a total system sampling rate constraint. In this paper, we specialize this broader goal to the setting of counting processes and to the context of tracking citation counts of researchers.   

References that are most closely related to our work are \cite{Kolobov19a} and \cite{Wang19a}. Reference \cite{Kolobov19a} considers the problem of finding optimal crawl rates to keep the information in a search engine fresh while maintaining the constraints on crawling rates imposed by the websites and also the total crawl rate constraint of the search engine, in the presence of non-uniform importance scores and change rates for the websites. Reference \cite{Wang19a} focuses on remote real-time reconstruction of a single Poisson counting process using uniform sampling. While taking more samples helps reconstruct the signal better, this increases queuing delays, which inherently affects the real-time signal estimation negatively. Reference \cite{Wang19a} studies this trade-off and finds the optimal sampling rate. Our timeliness metric is similar to the one considered in \cite{Wang19a}, which is the difference of a counting process and its sampled (updated, in our case) version. However, we consider multiple counting processes with different arrival rates and importance factors, and optimize our update rates for all processes jointly under a total update rate constraint. Similar to \cite{Kolobov19a}, we consider exponential arrival and sampling times in a constrained manner, and allow for importance coefficients, however, our timeliness metric and our overall problem setting are different.

In this paper, we first find an analytical expression for the long-term average difference between the actual and updated counting processes. We then minimize this expression as a function of the update rates of individual researchers subject to the overall update rate. We show that the optimal update rates are proportional to the \emph{square roots} of mean citation rates of the researchers for constant importance factors. Thus, it is optimal to update more prolific researchers more often, however, the proportionality is sub-linear and in the form of square root of the mean citation rate, i.e., there are diminishing returns due to the concavity (sub-linearity) of the square root function. We show that if the importance factors of the researchers are linear in their mean citation rates, then the optimal update rates are linear in their mean citation rates as well. We finally remark that other \emph{square root} results have appeared in completely different settings in caching problems. In particular, it was found in \cite{Yates17b} that, in order to keep the files fresh in a caching system, the (uniform) update rates of the files should be chosen proportional to the square roots of their popularity indices. In addition, it was shown in \cite{Cohen02} that, in order to minimize the average number of websites searched, the number of websites that a file is mirrored should be chosen proportional to the square root of the file's popularity.

\section{System Model and Problem Formulation} \label{sect:system_model}
\begin{figure}[t]
	\centerline{\includegraphics[width=0.9\columnwidth]{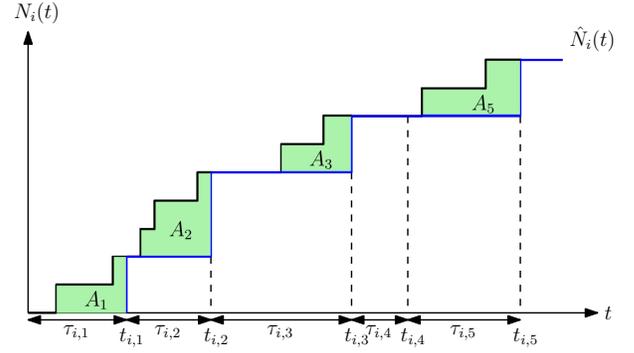}}
	\caption{The number of citations, $N_i(t)$, and the estimated number of citations, $\hat{N}_i(t)$, for researcher $i$. $A_j$ denotes the total estimation error in $[t_{i,j-1}, t_{i,j})$.}
	\label{Fig:dist_eval}
	\vspace{-0.4cm}
\end{figure}

Consider $n$ researchers. Let $N_i(t)$ denote the number of citations of researcher $i$. We model $N_i(t)$ as a Poisson process with rate $\lambda_i$, and assume that $N_i(t)$, for $i=1,\dots,n$, are independent. Let $t_{i,j}$ denote the time instance when the updater updates the number of citations of researcher $i$ for the $j$th time. We denote the inter-update time between the $j$th and $(j-1)$th updates for researcher $i$ as $\tau_{i,j}$. Based on these samples, the updater generates a real-time estimate of the counting process $N_i(t)$ as $\hat{N}_i(t)$, where
\begin{align}
\hat{N}_i(t) = N_i(t_{i,j-1}), \quad t_{i,j-1} \leq t< t_{i,j}.
\end{align}               
Fig.~\ref{Fig:dist_eval} shows $N_i(t)$ and $\hat{N}_i(t)$ with black and blue lines.

Similar to \cite{Wang19a}, we use the average difference between the actual and updated processes as a measure of timeliness, 
\begin{align}
\Delta_i(T) = \frac{1}{T}\int_{0}^{T} \left(N_i(t)- \hat{N}_i(t)\right) dt.\label{eqn:ave_dist}
\end{align}
If there are $m$ updates in the interval $[0,T]$, then $\Delta_i(T)$ is
\begin{align}\label{ave_dist}
\Delta_i(T) = \frac{1}{T}\sum_{j=1}^{m} A_j,
\end{align} 
where $A_j$ is the difference in the interval $[t_{i,j-1},t_{i,j} )$, see Fig.~\ref{Fig:dist_eval}. Then, the long term average difference for researcher $i$ is $\Delta_i=\lim_{T\rightarrow\infty} \Delta_i(T)$, and can be written as \cite{Najm17}, 
\begin{align}\label{ave_dist1}
\Delta_i = \lim\limits_{T\rightarrow \infty} \frac{1}{T}\sum_{j=1}^{m} A_j  
= \lim\limits_{T\rightarrow \infty} \frac{m}{T} \cdot \frac{1}{m} \sum_{j=1}^{m} A_j  = \rho_i \mathbb{E}[A].
\end{align}

Similar to the derivation in \cite{Wang19a}, conditioned on an arbitrary $j$th inter-update time, i.e., $\tau_{i} =\tau_{i,j}  = d$, and the number of citation arrivals in that time interval $[t_{i,j-1},t_{i,j-1}+d)$, i.e., $\tilde{N}_i(d) = N_i(t_{i,j-1}+d)-N_i(t_{i,j-1}) =k$, the expected difference is 
\begin{align}
 \mathbb{E}\left[A|\tilde{N}_i(d) = k, \tau_{i} = d\right] = \frac{kd}{2}.   
\end{align}
Thus, we have 
\begin{align}\label{eqn:cond_exp}
 \mathbb{E}\left[A| \tau_{i} = d\right] = \mathbb{E}\left[\mathbb{E}\left[A|\tilde{N}_i(d) = k, \tau_{i} = d\right]\right] 
 = \frac{\lambda_i d^2}{2}.
\end{align}
In the following subsections, we present three different models for updating the citation numbers.  

\subsection{Model 1: Poisson Updater} \label{subsect:model1}
\begin{figure}[t]
	\centering  \includegraphics[width=0.90\columnwidth]{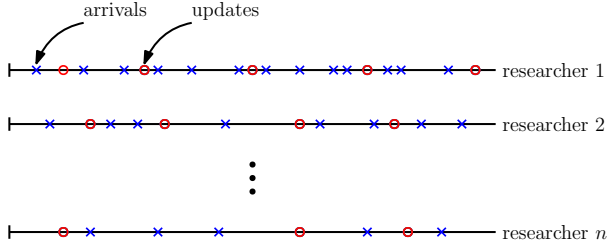}
	\caption{Poisson updater: Inter-update times are exponential with rate $\rho_i$.}
	\label{fig:model1}
\end{figure}

In this model, shown in Fig.~\ref{fig:model1}, the inter-update times for researcher $i$ are exponential with rate $ \rho_i$. Update processes for different researchers are independent. Continuing from (\ref{eqn:cond_exp}), we find $\mathbb{E}[A]$ using exponential distribution as,    
\begin{align} \label{eqn_exp_a}
\mathbb{E}[A] =\int_{0}^{\infty}\mathbb{E}\left[A| \tau_{i} = t\right] f_{\tau_i}(t)dt 
= \frac{\lambda_i}{2}\mathbb{E}\left[\tau_{i}^2\right] = \frac{\lambda_i}{\rho_i^2}. 
\end{align}
Thus, the long term average difference $\Delta_i$ in (\ref{ave_dist1}) with a Poisson updater is 
\begin{align} \label{poisson-updater-result}
\Delta_i = \frac{\lambda_i}{\rho_i}.
\end{align}

\subsection{Model 2: Deterministic Updater}  \label{subsect:model2}
\begin{figure}[t]
	\centering  \includegraphics[width=0.90\columnwidth]{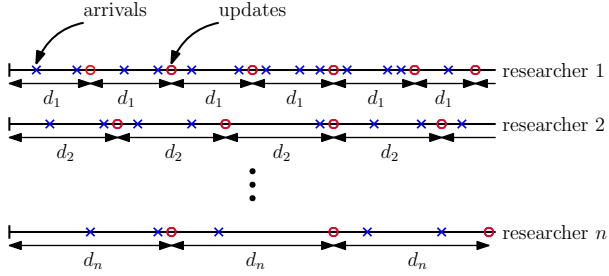}
	\caption{Deterministic updater: Inter-update times are equal with $ d_i = \frac{1}{\rho_i}$.}
	\label{fig:model2}
	\vspace{-0.4cm}
\end{figure}

In this model, shown in Fig.~\ref{fig:model2}, the inter-update times are deterministic and chosen optimally. Similar to \cite{Yates17b}, given that there are $m_i$ updates for researcher $i$ in the time interval $[0,T]$, the optimal inter-update times should be chosen equal to each other, i.e., $\tau_{i,j} = \frac{T}{m_i+1}$, for all $j$. Letting $T\rightarrow \infty$, this update scheme results in uniform sampling with rate $\rho_i$ for researcher $i$ where $\rho_i  = \lim_{T\rightarrow \infty}\frac{m_i}{T}$. By using $d_i =\frac{1}{\rho_i}  $ and (\ref{eqn:cond_exp}), we obtain $\mathbb{E}[A] = \frac{\lambda_i}{2\rho_i^2}$. 
Thus, the long term average difference $\Delta_i$ in (\ref{ave_dist1}) with a deterministic (and uniform) updater is 
\begin{align} \label{det-updater-result}
\Delta_i = \frac{\lambda_i}{2\rho_i}.
\end{align}

\subsection{Model 3: Common Synchronized Probabilistic Updater }  \label{subsect:model3}
\begin{figure}[t]
	\centering  \includegraphics[width=0.90\columnwidth]{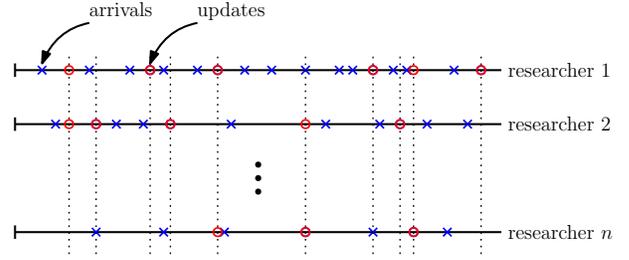}
	\caption{Common synchronized updater: Common synchronized inter-update times are exponential with rate $\rho$. At each common update opportunity, researcher $i$ is updated with probability $p_i$}.
	\label{fig:model3}
	\vspace{-0.4cm}
\end{figure}

In this model, shown in Fig.~\ref{fig:model3}, the updater has a common synchronized update schedule that applies to all researchers. The inter-update times of the common updater are exponential with rate $\rho$. At each update instant, researcher $i$ is updated with probability $p_i$ independently of other researchers. Thus, inter-update times for researcher $i$ are exponential with rate $\rho p_i$. Note that here, we create the Poisson updates for researcher $i$ by thinning the Poisson common updates using probabilistic updates according to $p_i$. The main problem, therefore, is to choose $p_i$ for each researcher as it determines its mean update rate. This problem is the same as the one in Section~\ref{subsect:model1} and in the optimal policy $p_i$ is $p_i = \frac{\rho_i}{\rho}$ assuming $\rho$ is sufficiently large to have feasible $p_i$, i.e., $0\leq p_i\leq 1$, for all $i$. 

\subsection{Problem Formulation}
Researcher $i$ has the mean citation rate $\lambda_i$. In addition, and optionally, we consider an importance factor, $\mu_i$, for researcher $i$. This may be removed by choosing all $\mu_i=\mu$. Then, the total long term average difference (over all researchers) becomes $\Delta = \sum_{i=1}^{n} \mu_i\Delta_i$, where per researcher difference, $\Delta_i$, is given by (\ref{poisson-updater-result}) for the Poisson updater and common synchronized updater models, and by (\ref{det-updater-result}) for the deterministic updater model. The expressions in  (\ref{poisson-updater-result}) and (\ref{det-updater-result}) differ only by a factor of 2, which is inconsequential for optimization purposes. Therefore, without loss of generality, from now on, we use the expression in (\ref{poisson-updater-result}). In addition, due to computational limitations, the updater is subject to a total update rate constraint  $\sum_{i=1}^{n}\rho_i\leq c$. Our aim is to find the optimal update rates for all researchers, $\rho_i$, for $i=1,\dots,n$, such that the total long term average difference, $\Delta$, is minimized while satisfying the constraint on the total update rate. Thus, our optimization problem is,
\begin{align}
\label{problem1_mod}
\min_{\{\rho_i \}}  \quad &  \sum_{i=1}^{n}\frac{\mu_i \lambda_i}{\rho_i} \nonumber \\
\mbox{s.t.} \quad & \sum_{i=1}^{n} \rho_i\leq c \nonumber \\
\quad & \rho_i\geq 0, \quad i=1,\dots,n.
\end{align} 
We solve the optimization problem in (\ref{problem1_mod}) in the next section.                        

\section{The Optimal Solution} \label{sect:opt_soln}
The optimization problem in (\ref{problem1_mod}) is convex as the cost function is convex and the constraints are linear. We introduce the Lagrangian function \cite{Boyd04} for (\ref{problem1_mod}) as
\begin{align}
\mathcal{L} = \sum_{i=1}^{n}\frac{\mu_i \lambda_i}{\rho_i} + \beta\left(\sum_{i=1}^{n} \rho_i- c\right)- \sum_{i=1}^{n}\nu_i\rho_i,
\end{align}
where $ \beta \geq 0$ and $\nu_i\geq 0$ for all $i$. Next, we write the KKT conditions as
\begin{align}\label{KKT}
\frac{\partial \mathcal{L}}{\partial \rho_i} = -\frac{\mu_i \lambda_i}{\rho_i^2} +\beta-\nu_i= 0,
\end{align}
for all $i$, and the complementary slackness conditions as 
\begin{align}
\beta\left(\sum_{i=1}^{n} \rho_i- c\right) &= 0, \label{CS_1}\\
\nu_i \rho_i &= 0, \label{CS_2}
\end{align}
for all $i$. Since the optimization problem in (\ref{problem1_mod}) is convex, the KKT conditions are necessary and sufficient.

First, we observe that the total update rate constraint $\sum_{i=1}^{n} \rho_i\leq c$ must be satisfied with equality. If there is an update rate allocation policy such that $ \sum_{i=1}^{n} \rho_i< c$, then we can achieve a lower average difference by increasing any $\rho_i$ as the cost function of (\ref{problem1_mod}) is a decreasing function of $\rho_i$. Thus, in the optimal update rate allocation policy, we must have $\sum_{i=1}^{n} \rho_i= c$ and $\beta\geq0$ due to (\ref{CS_1}). 

Next, we note that in the optimal policy, we must have $ \rho_i>0$, for all $i$, as $\rho_i = 0$ leads to infinite objective function in (\ref{problem1_mod}) which clearly cannot be an optimal solution. Thus, for the optimal rate allocations, we have $\rho_i >0$ and $\nu_i = 0$, for all $i$, due to (\ref{CS_2}). 

From (\ref{KKT}), we find $\rho_i = \sqrt{\frac{\mu_i \lambda_i}{\beta}}$. By using $ \sum_{i=1}^{n} \rho_i= c$, we solve $\beta = \frac{\left(\sum_{i=1}^{n} \sqrt{\mu_i \lambda_i}\right)^2}{c^2}$, which gives the optimal policy,
\begin{align}
\rho_i =  \frac{c\sqrt{\mu_i \lambda_i}}{\left(\sum_{j=1}^{n}\sqrt{\mu_j \lambda_j}\right)}, \quad i=1,\dots,n. \label{opt_policy}
\end{align}   
Using the optimal rate allocation policy in (\ref{opt_policy}), we obtain
\begin{align}
\Delta_i = \frac{\sqrt{\lambda_i}\left(\sum_{j=1}^{n}\sqrt{\mu_j \lambda_j}\right)}{c\sqrt{\mu_i}}, \quad i=1,\dots,n, \label{opt_dist}
\end{align}
and the total long term average difference $\Delta$ as
\begin{align}
\Delta = \frac{\left(\sum_{j=1}^{n}\sqrt{\mu_j \lambda_j}\right)^2}{c}.\label{total_opt_dist}
\end{align}

Thus, the optimal update rates allocated to researchers in (\ref{opt_policy}) are proportional to the square roots of their importance factors, $\mu_i$, multiplied by their mean citation rates, $\lambda_i$. We note that if we ignore the importance factors, i.e., $\mu_i=\mu=1$, then the optimal update rates are proportional to the square roots of the mean citation rates. On the other hand, if we choose the importance factors as proportional to the mean citation rates, i.e., $\mu_i=\alpha \lambda_i$, then the optimal update rates become linear in the mean citation rates.
                             
\section{Numerical Results} \label{sect:num_res}
In this section, we provide three numerical results. In the first two examples, we choose the mean citation rates as
\begin{align}
\lambda_i = a r^{i}, \quad i =1,\dots, n, \label{arrival_rates}
\end{align}
where $a> 0$ and $0< r\leq 1$.

\begin{figure}[t]
	\begin{center}
		\subfigure[]{%
			\includegraphics[width=0.69\linewidth]{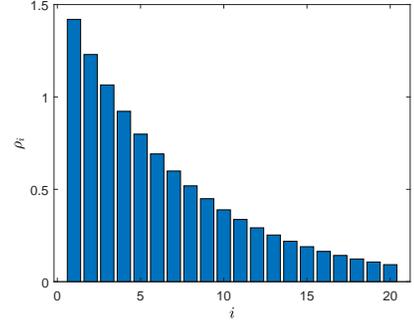}}\\
		\subfigure[]{%
			\includegraphics[width=0.69\linewidth]{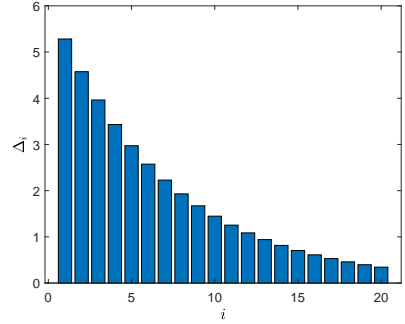}}
	\end{center}
	\caption{(a) Optimal update rate allocation for each researcher, and (b) the corresponding optimal long term average difference $\Delta_i$, when we use uniform importance coefficients $\mu_i=1$, with $\lambda_i$ given in (\ref{arrival_rates}), with $a =10$ and $r = 0.75$ for $n=20$.}
	\label{Fig:sim1}
	\vspace{-0.4cm}
\end{figure}

In the first example, we take $a=10$, $r=0.75$, $n=20$ and $c=10$. For this example, we use uniform importance coefficients, i.e., $\mu_i = 1$, for all $i$. We observe in Fig.~\ref{Fig:sim1}(b) that researchers with higher mean citation rates have higher long term average difference $\Delta_i$ even though they are updated with higher update rates shown in Fig.~\ref{Fig:sim1}(a). Further, we observe in Fig.~\ref{Fig:sim1}(a) that due to diminishing returns caused by the square root allocation policy, update rates of the researchers with low mean citation rates are still comparable to the update rates of the researchers with high mean citation rates.

\begin{figure}[t]
	\begin{center}
		\subfigure[]{%
			\includegraphics[width=0.69\linewidth]{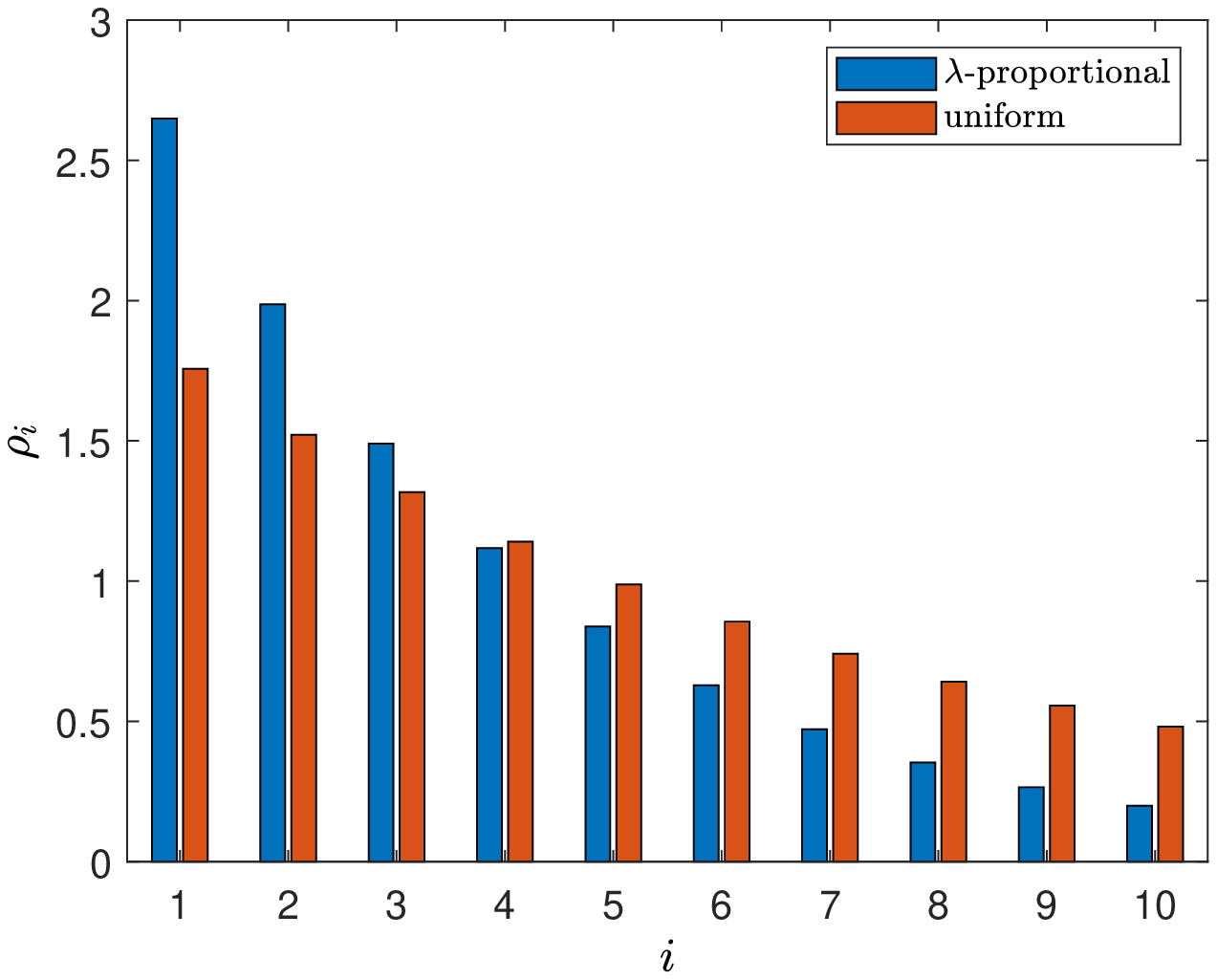}}\\
		\subfigure[]{%
			\includegraphics[width=0.69\linewidth]{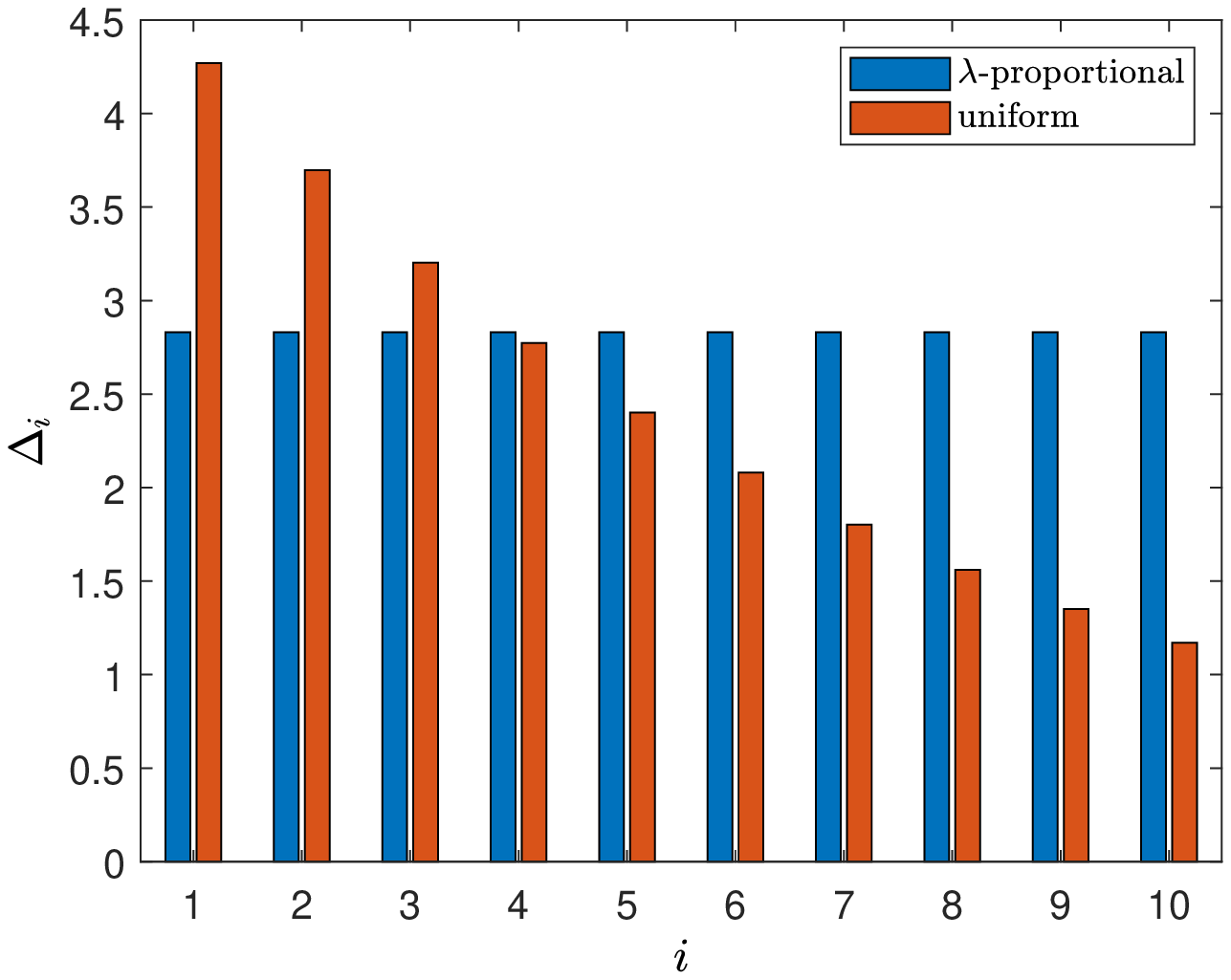}}
	\end{center}
	\caption{(a) Optimal update rate allocation for each researcher, and (b) the corresponding optimal long term average difference $\Delta_i$, when we use \emph{$\lambda$-proportional} and uniform importance coefficients, with $\lambda_i$ given in (\ref{arrival_rates}), with $a =10$ and $r = 0.75$ for $n=10$.}
	\label{Fig:sim2}
	\vspace{-0.4cm}
\end{figure}

In the second example, we consider the case where the importance factors are chosen proportional to the mean citation rates of the researchers. We call such coefficients as \emph{$\lambda$-proportional} importance coefficients, which are given by
\begin{align}
    \mu_i = \frac{\lambda_i}{\sum_{j=1}^{n}\lambda_j},\quad i=1,\dots,n.
\end{align}
In order to make a fair comparison between the \emph{$\lambda$-proportional} and \emph{uniform} importance coefficients, we scale the uniform importance coefficients as $\mu_i = \frac{1}{n}$, for all $i$.

As mentioned at the end of Section~\ref{sect:opt_soln}, when we use \emph{$\lambda$-proportional} importance coefficients, the optimal update rates become linear in the mean citation rates. We observe in Fig.~\ref{Fig:sim2}(a) that using \emph{$\lambda$-proportional} importance coefficients favor researchers with higher mean citation rates as their update rates increase compared to the update rates with the uniform importance coefficients. Further, we observe in Fig.~\ref{Fig:sim2}(b) that the long term average differences are equal to each other when \emph{$\lambda$-proportional} importance coefficients are used. 

\begin{figure}[t]
	\centerline{\includegraphics[width=0.85\columnwidth]{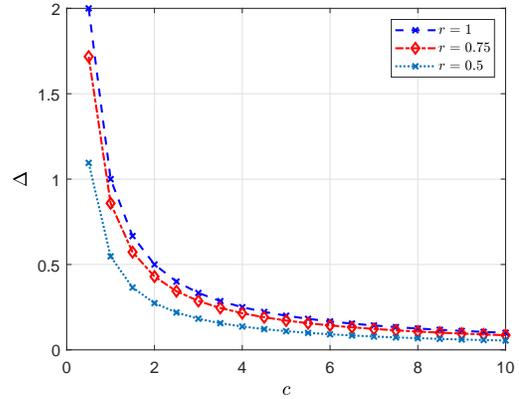}}
	\caption{Total long term average difference $\Delta$ with respect to $c$, when uniform importance coefficients are used and $\lambda_i$ are given in (\ref{alt_dist}), with $a=1$ and $r= 0.5,0.75, 1$ for $n= 10$. }
	\label{Fig:sim3}
	\vspace{-0.4cm}
\end{figure}

In the third example, we choose the mean citation rates as
\begin{align}
    \lambda_i = \frac{a r^i}{\sum_{j=1}^n r^j}, \quad i =1,\dots,n, \label{alt_dist}
\end{align}
which satisfy $ \sum_{i=1}^n \lambda_i=a$. Note that, by this selection, we force total citation means of all researchers to be a constant. For this example, we take $n= 10$, $a =1$ and consider three different $r$, which are $r =0.5,0.75, 1$.  Note also that, a smaller $r$ corresponds to a less even (more polarized) distribution of total mean citation rates among the researchers. We use uniform importance coefficients and plot achieved $\Delta$ with respect to $c$ in Fig.~\ref{Fig:sim3}. We observe in Fig.~\ref{Fig:sim3} that more polarized distribution of mean citation rates (smaller $r$) yields a lower $\Delta$ for the system, as we exploit the differences among the researchers by allocating even higher update rates to researchers with higher mean citation rates. As an aside, we note that if we used \emph{$\lambda$-proportional} importance coefficients, we would have a $\Delta$ which is independent of individual $\lambda_i$ that depends only on the sum of $\lambda_i$ which is $a$ here. This achieved $\Delta$ is also equal to the $\Delta$ achieved with uniform importance coefficients when $r= 1$ which is shown as the blue dashed line in Fig.~\ref{Fig:sim3}. Thus, if we use \emph{$\lambda$-proportional} importance coefficients, the achieved $\Delta$ is independent of the mean citation rate distribution among the researchers, but it results in higher $\Delta$. In other words, uniform importance coefficients achieve lower $\Delta$ compared to \emph{$\lambda$-proportional} importance coefficients in this case for $r<1$.           

\section{Conclusion, Discussion and Future Directions} \label{sect:dis}     
We considered the problem of timely updating of citation counts by a resource-constrained updater. We showed that the optimal policy is to choose the update rates of individual researchers proportional to the square roots of their mean citation rates multiplied by their importance factors (if any). 

Next, we discuss limitations of our model and suggest future research directions. First, we note that we modeled the citation numbers of a researcher as a counting process, which is monotonically increasing and increments one at a time. The monotonically increasing nature of the counting process assumes that published articles are always available and cannot be withdrawn or changed, which may not be the case, especially if the crawler cannot reach certain websites that it used to reach, and lose access to previously counted publications. In addition, one at a time increments assume that citations come one by one, which may not be the case as conferences publish all the articles simultaneously in proceedings, and journals publish articles in monthly issues. That is, the increments in citations for each researcher may be more than one at a time. Thus, modeling the number of citations with a simple Poisson process may not be sufficient. However, as the researchers increasingly publish their works as they are completed on their personal websites or academic websites such as arXiv, it may still be acceptable to model arrivals as Poisson processes. Even then though, multiple citations to the same researcher from a single article will not be captured by the model in this paper. In addition, considering the fact that researchers often collaborate and publish articles jointly, the arrivals of citations for different researchers might be correlated. In this paper, we considered the case where the citation arrivals for researchers are independent counting processes. This may be feasible by focusing on researchers with no common publications.

Finally, we assumed that mean citation rates and importance factors are fixed and known by the updater. As a future direction, an online setting can be considered where both of these parameters are learned from observations over time. 

\bibliographystyle{unsrt}
\bibliography{IEEEabrv,lib_v1_melih}
\end{document}